\newcommand*\diff{\mathop{}\!\mathrm{d}}

\documentclass{IEEEtran4PSCC}
\usepackage[svgnames]{xcolor}
\usepackage{booktabs}% http://ctan.org/pkg/booktabs
\usepackage{tabularx}% http://ctan.org/pkg/tabularx
\usepackage{mathtools}
\usepackage{ifpdf}

\usepackage{cite}
\usepackage{braket}

\ifCLASSINFOpdf
   \usepackage[pdftex]{graphicx}
\else
   \usepackage[dvips]{graphicx}
\fi
\usepackage{algorithm,algorithmic}
\usepackage{amsfonts}
\usepackage{quantikz}
\usepackage{tikz}
\usetikzlibrary{arrows.meta,shapes,positioning,fit,backgrounds,calc}
\usepackage{amsmath}
\makeatletter
\let\old@ps@headings\ps@headings
\let\old@ps@IEEEtitlepagestyle\ps@IEEEtitlepagestyle
\def\psccfooter#1{%
    \def\ps@headings{%
        \old@ps@headings%
        \def\@oddfoot{\strut\hfill#1\hfill\strut}%
        \def\@evenfoot{\strut\hfill#1\hfill\strut}%
    }%
    \def\ps@IEEEtitlepagestyle{%
        \old@ps@IEEEtitlepagestyle%
        \def\@oddfoot{\strut\hfill#1\hfill\strut}%
        \def\@evenfoot{\strut\hfill#1\hfill\strut}%
    }%
    \ps@headings%
}
\makeatother

\psccfooter{%
        \parbox{\textwidth}{\hrulefill \\ \small{24th Power Systems Computation Conference} \hfill \begin{minipage}{0.2\textwidth}\centering \vspace*{4pt} \includegraphics[scale=0.06]{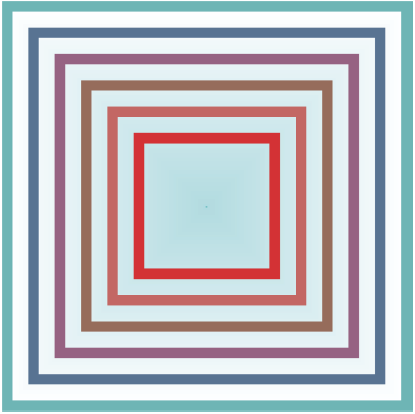}\\\small{PSCC 2026} \end{minipage} \hfill \small{Limassol, Cyprus --- June 8-12, 2026}}%
}

\begin{document}
\renewcommand{\ttdefault}{cmtt}
\bstctlcite{IEEEexample:BSTcontrol}
% \title{Quantum Deep Reinforcement Learning-based Dynamic Security Control in IBR-Rich Grids}

\title{Quantum-Embedded Dynamic Security Control using Hybrid Deep Reinforcement Learning}

%% To specify the authors when (number of affiliations <= 2)
% \author{
% \IEEEauthorblockN{Author n.1 Name per Affiliation A\\ Author n.2 Name per Affiliation A}
% \IEEEauthorblockA{Department of Electrical Engineering and Computer Science \\
% Name of the organization, acronyms acceptable\\
% City, Country\\
% \{email author n.1, email author n.2\}@domain (if desired)}
% \and
% \IEEEauthorblockN{Author n.1 Name per Affiliation B\\ Author n.2 Name per Affiliation B}
% \IEEEauthorblockA{(Affiliation B) Department Name of Organization \\
% Name of the organization, acronyms acceptable\\
% City, Country\\
% \{email author n.1, email author n.2\}@domain (if desired)}
% }

\author{\IEEEauthorblockN{Amin Masoumi and Mert Korkali}
\IEEEauthorblockA{\textit{Department of Electrical Engineering and Computer Science}\\ 
\textit{University of Missouri} \\
Columbia, MO 65211 USA \\
e-mail: \{\texttt{am4n5,korkalim\}@missouri.edu}}}

%% To specify the authors when (number of affiliations > 2)
% \author{\IEEEauthorblockN{Author n.1\IEEEauthorrefmark{1},
% Author n.2\IEEEauthorrefmark{2},
% Author n.3\IEEEauthorrefmark{3}, 
% Author n.4\IEEEauthorrefmark{3} and
% Author n.5\IEEEauthorrefmark{4}}
% \IEEEauthorblockA{\IEEEauthorrefmark{1} Department Name of Organization A\\
% Name of the organization A,
% Address A\\ Emails if wanted}
% \IEEEauthorblockA{\IEEEauthorrefmark{2} Department Name of Organization B\\
% Name of the organization B,
% Address B\\ Emails if wanted}
% \IEEEauthorblockA{\IEEEauthorrefmark{3} Department Name of Organization C\\
% Name of the organization C,
% Address C\\ Emails if wanted}
% \IEEEauthorblockA{\IEEEauthorrefmark{4}Department Name of Organization D\\
% Name of the organization D,
% Address D\\ Emails if wanted}
% }

\maketitle

\begin{abstract}
Dynamic security control (DSC) is considered a pivotal step for the future power grid, which is increasingly penetrated by inverter-based resources. However, the efficiency of such practices, whether governed by automatic generation control or virtual inertia scheduling, can be intractable due to the complexity of the problem and the need to solve the differential-algebraic equation in a timely manner with the required accuracy. In this regard, the model-free deep reinforcement learning algorithm demonstrates reliable performance. In addition, the introduction of fault-tolerant and near-term quantum computing terminologies, i.e., noisy intermediate-scale quantum, opens avenues for improving the performance of model-free algorithms leveraging quantum capabilities. This paper provides an organized framework and assesses its dependability by evaluating the performance of a quantum-embedded algorithm on the DSC of the IEEE 39-bus test system. Hence, the obtained results demonstrate promising applications, along with shortcomings that can be addressed and further developed later.
\end{abstract}


% Use this to place sponsorships
\thanksto{\noindent Submitted to the 24th Power Systems Computation Conference (PSCC 2026).}

\section{Introduction and Background}

The widespread deployment of inverter-based resources
(IBRs), such as solar photovoltaics and wind power, has positive impacts on key areas, including decarbonization, modernization, lower long-term energy costs, and industry growth~\cite{bevrani2017microgrid}. This transition can be traced to the introduction of significant challenges to the dynamic security control of power grids (DSC) arising from the displacement of traditional synchronous generators (SGs)~\cite{shazon2022frequency}. In this regard, the pivotal challenge is the loss of physical inertia, which is critically essential for maintaining grid stability during severe contingencies, such as three-phase short-circuit faults~\cite{wang2025transient}, particularly at intersections with complex power flow. By diving deeper, it can be observed that SGs inherently provide inertia to the power system through the kinetic energy stored in their physical rotating masses. This natural inertia induces a turbine-governor-based frequency response to counteract rapid changes during a fault and extends the critical clearing time (CCT)~\cite{jafari2024role}. This can effectively slow down frequency deviations by absorbing or distributing transient energy to other actors, i.e., SGs. Hence, the out-of-step condition (OOS) can be prevented from spreading to other intersections and triggering cascading failure. Specifically, inherent inertia limits the acceleration range and provides a manageable deceleration range, thereby improving dynamic security, i.e., transient stability. In contrast, IBRs must be connected to their power electronic converters, grid-following (GFL) or grid-forming (GFM) \cite{fang2025dynamic}, to operate in the power grid. This lack of physicality in mass leads to curtailment of inertia~\cite{saleem2024assessment}. Furthermore, advanced control strategies~\cite{zhang2021grid} have been developed to emulate the inertial response of SGs, leading to the concept of a virtual synchronous generator (VSG). However, the VSG can put constraints on the converter’s bandwidth and energy storage capacity~\cite{abdelghany2025enhanced}. Hence, the probable repercussion is a delayed response compared to the turbine-governor-based counterparts. Moreover, the deceleration energy is diminished during the noted severe contingency. Hence, the shrinking deceleration area reduces the CCT, making IBR-rich grids~\cite{saha2023impact} prone to cascading failures and major blackouts. In other words, this makes the IBR-rich grids extremely weak at responding to frequency stress, i.e., sudden load increases. This arises from concerns regarding the performance of these converters during islanding events, such as the disconnection of a bus or transmission line. The creation of sophisticated control methods becomes increasingly vital in ensuring DSC of IBRs. In this regard, the virtual inertia and reference power~\cite{ma2025coordinated,chen2021enhanced} play a key role in balancing the deceleration and acceleration rates. However, this perspective requires a control system that is sufficiently aware of state transitions. Among various disciplines, model-free deep reinforcement learning (DRL) techniques have demonstrated scalable performance for DSC under partial observations and varying operating conditions. Unlike traditional control methods that rely solely on exact models, model-free DRL agents can learn optimal control policies directly from interactions with the environment and train their policies based on experience. This capability is remarkably vital for estimating the trajectories of differential-algebraic equations (DAEs) in complex, nonlinear, and high-dimensional environments, such as IBR-rich grids.

% The authors in~\cite{chen2023investigation} have used deep deterministic policy gradients with a softmax function (SD2) to control the performance of VSG under severe transient instability conditions. This model-free technique is based on the DDPG's functionality, leveraging online actor and critic capabilities. In this regard, the SD2 agent attempts to restore post-fault rotor stability by adjusting the virtual inertia and the reference active power. Hence, the reward function penalizes instability and angular-frequency deviation in the system. However, the speed of parameter optimization, i.e., the gradient calculation for the weight and bias, is constrained by policy exploration, which simultaneously affects the accuracy of the actor and critic. As the environment becomes more complex, convergence can take longer.

In \cite{eskandari2023deep,oboreh2023virtual,lee2024deep,chen2023investigation}, deep deterministic policy gradient (DDPG), twin delayed deep deterministic policy gradient (TD3), and softmax DDPG have been developed to control a GFM converter. The main role of the conducted DDPG is to emulate virtual inertia through impedance sharing under time-varying load conditions. The target of the reward function is to increase the frequency response of IBR-rich grid regarding the rate of change of frequency (RoCoF), angular frequency, etc. In \cite{yang2022distributed} and ~\cite{eskandari2023convolutional} hybrid policy- and value-based DRL method, soft actor-critic (SAC), has been applied to the control operation of VSGs. This continuous-action-space algorithm aims to suppress power oscillations by regulating the frequency response in a varying environment with partial observability. Hence, the reward function comprises three terms: frequency deviation, inertia parameters, and droop gains. However, partial observability is a hurdle to the coordination of multiple agents, underscoring the need for a framework that can handle the environment's nonlinearity and high complexity. In~\cite{ma2024dynamic}, meta-RL has been brought into the effect of the power-frequency droop controller regarding fast frequency regulation. The model-free DRL has been organized in a hierarchical architecture to adapt to sporadic operating conditions, targeting both quantitative and qualitative factors, including ancillary services, cost, and frequency response. However, the training process is susceptible to stability issues during frequency nadir and peak times. It is evident that each continuous-action-based DRL methodology encounters significant challenges, potentially affecting its feasibility in real-world applications~\cite{lockwood2020reinforcement,chen2024deep}. The high-dimensionality of state and action spaces can create scalability issues that make it difficult for the specific agent to explore a variety of policies and converge with the granularity of the solutions. Furthermore, sample inefficiency is a common bottleneck for every model-free technique. In this regard, interactions with the power system environment can be resource-intensive due to the time-consuming process of solving DAEs and the storage of data. In other words, there are inherent limitations that require tailoring case studies to the ever-changing dynamics of contingencies, e.g., system loading, fault location, and fault duration. Consequently, the tradeoff between exploration and exploitation can be adversely affected, leading to suboptimal performance. Quantum mechanics can address these obstacles by providing tools that can remediate complexity. The main philosophy of using a quantum perspective is to leverage its properties (i.e., superposition and entanglement) to facilitate the DRL agent's learning. The key notion here is the application of ansatz or a parameterized quantum circuit (PQC)~\cite{wu2025quantum} that can learn the interaction of computing units, e.g., actor and critic, for the benefit of accuracy, along with convergence speed.
In this paper, the primary focus is to leverage PQC to boost the DRL agent's capability with fewer trainable parameters. Therefore, we have devised a unique ansatz that enables exploration of the continuous action space of VSG parameters. The proposed strategy aims to alleviate the complexity of high-dimensional state spaces by stimulating quantum-mechanical dynamics and fitting the behavior of excited states to estimate the impact of changing the VSG parameter at each time step. Hence, our proposed algorithm is the first of its kind to improve DSC of IBR-rich grid by increasing the DRL agent's adaptivity via integrating PQC to increase the processing of highly nonlinear relationships between states and actions. 

The rest of the paper is organized as follows: Section II describes the problem formulation for dynamic security control of an IBR-rich grid. Section III describes the architecture behind the proposed quantum DRL. Sections IV and V define the case study's implementation scope and the results obtained. Finally, Section VI concludes the paper. 

\section{Problem Formulation}
As mentioned above, the absence of natural inertia in IBR-rich grids necessitates the application of control algorithms for emulation, i.e., VSG. In addition, restoring the system frequency to its nominal value is the main objective of DSC. Hence, the conventional frequency response, governed by automatic generation control, must integrate wide-area actions from the internal setpoints of the VSGs and virtual inertia.  This is because severe contingencies that violate system synchronism replace the core objective of DSC with stability of the transient angle as the dominant concern. This can be inferred from the probable cascading failures and the spread of blackouts that lead to system collapse. In other words, correcting the power imbalance in the stressed post-fault state can exacerbate rotor angle swings and spread the instability to other areas. Hence, the DSC objective must be formulated to address dynamic security controls in such an environment. Specifically, performance and stability are the targets here:
\begin{subequations}\label{eq:eq2block}
\begin{align}
% \min_{\Delta \omega(t),\, \frac{d\omega}{dt},\, \delta_{\max}(t)} \quad
% & \int_{t_0}^{t_f} \!\Big[ \alpha_1 \Delta \omega(t)^2 + \alpha_2 \!\left(\frac{d\omega}{dt}\right)^{\!2} \Big]
%   + \beta \max_{t \in [t_0,t_f]} \!\big|\delta_{\max}(t) - \delta_{\mathrm{COI}}(t)\big| \; \diff t
\begin{split}
\min_{\Delta \omega(t), \frac{\diff \omega}{dt}, \delta_{\max}(t)} & \quad \int_{t_0}^{t_f} \left[ \xi_1 \Delta \omega(t)^2 + \xi_2 \left(\frac{\diff \omega}{\diff t}\right)^2\right] \\
& + \left[\zeta \cdot \max_{t \in [t_0, t_f]} |\delta_{\max}(t) - \delta_{\text{COI}}(t)| \right]  \diff t
\end{split}
\label{eq:eq1}\\
\intertext{\text{s.t.}}
J_i(t)\,\frac{\diff \omega_i}{\diff t} &=
P_{\mathrm{ref},i}(t) - P_{e,i}(t) - D_i(t)\,(\omega_i-\omega_n),
\label{eq:eq2}\\
P_{e,i} &= V_i \sum_{j=1}^{N_b} V_j \big(G_{ij}\cos\delta_{ij} + B_{ij}\sin\delta_{ij}\big),
\label{eq:eq5}\\
\left|\frac{\diff \delta_i}{\diff t}\right| &\le \omega_{\max}^{\mathrm{dev}},
\label{eq:eq7}\\
\int_{t_0}^{t_f} \!\big|P_{e,i} - P_{\mathrm{ref},i}\big| \,\diff t &\le E_{\max},
\label{eq:eq8}\\
P_{\mathrm{ref}}^{\min} \;\le\; P_{\mathrm{ref},i}(t) &\le P_{\mathrm{ref}}^{\max},
\label{eq:eqf1}\\
J^{\min} \;\le\; J_i(t) &\le J^{\max},
\label{eq:eqf2}\\
D^{\min} \;\le\; D_i(t) &\le D^{\max},
\label{eq:eqf3}\\
\left|\frac{\diff P_{\mathrm{ref},i}}{\diff t}\right| &\le R_{\max}.
\label{eq:eqf4}
\end{align}
\end{subequations}

\noindent According to \eqref{eq:eq1}, the first and the second brackets are the performance and stability functions, respectively. In \eqref{eq:eq1}, $P_{\text{ref}}(t)$, $J(t)$, $D(t)$, $\Delta \omega(t)$, $\frac{\diff \omega}{\diff t}$, $\delta_{\max}(t)$, and $\delta_{\text{COI}}(t)$ are the reference active power, virtual inertia, virtual damping, angular frequency, RoCoF, maximum rotor angle, and center of inertia (COI) rotor angle, respectively. Also, $t_0$, $t_f$, $\xi_1 $, $\xi_2 $, and $\zeta$ are fault clearance, time of transient instability, and coefficients of DSC and transient stability, respectively. Moreover,~\eqref{eq:eq2} provides insight on the $i$th VSG’s swing equation~\cite{li2025power} to represent control constraints. Equation (\ref{eq:eq5}) provides grid's physics and operating points feasibility of actions with respect to the physics and operating points of the grid. Equation \eqref{eq:eq7} is formulated to monitor the deviations of the rotor angles. Equations \eqref{eq:eq8}--\eqref{eq:eqf4} are efficiency constraints with respect to the action space and physical limitations of the IBR-rich grid. As can be seen, the key perspectives of the presented formulation, as a hybrid optimal action problem, are continuous dynamics (VSG equations) and discrete events (severe contingency). Hence, we are dealing with a high-dimensional function space, along with non-convexity arising from power flow equations and stability constraints that aim to maintain continuous actions in response to discrete instability events. In addition, the complexity of the problem is magnified by the impact of $|\delta_{\max}(t)|$ on the performance of optimal actions. Therefore, the key challenge here is the highly nonlinear dynamics of actions that depend on the complex, evolving system, making the problem intractable for conventional real-time optimization techniques. This serves as an ideal setting for the implementation of an intelligent and learning-based methodology. This technique is capable of being implicitly conditioned to discern the relationship between control actions and transient stability limits.

% \subsection{Dynamic Security via Transient Stability}

% \subsection{Special Case of IBR-Rich Grids}

\section{Dynamic Security Control via  QDRL}
The post-fault trajectories, e.g., rotor angle, of interactions between system states and actions become increasingly untenable as in the case of capturing targets of frequency control and transient stability. This limitation calls for a smart algorithm, i.e., DRL, that can address the computational burden posed by the complex dynamics of IBRs (GFL or GFM).  However, model-free DRL faces two crucial challenges that hamper its performance: reactive responses and sample inefficiency. The reactive nature of DRL awaits the instability to develop over several control steps. In other words, the DRL agent chooses the best action for the current state and does not consider the best action to prevent instability predicted to occur, e.g., 150 ms after the fault clearance. This is because rapidly developing transient instability can become irreversible after separation of the rotor angles. More specifically, the lack of proactive response, i.e., waiting several control steps of observing the instability may defeat the feasibility of corrective actions even with excessively large values. On the other hand, regarding sample inefficiency, model-free DRL requires a large number of episodes ($500{,}000$-$10/100$ of millions) of interactions with the environment to learn semi-optimal policies. This is simply computationally expensive under a variety of contingencies and grid topologies. 

\begin{figure}[htbp]
\centerline{\includegraphics[width=\linewidth]{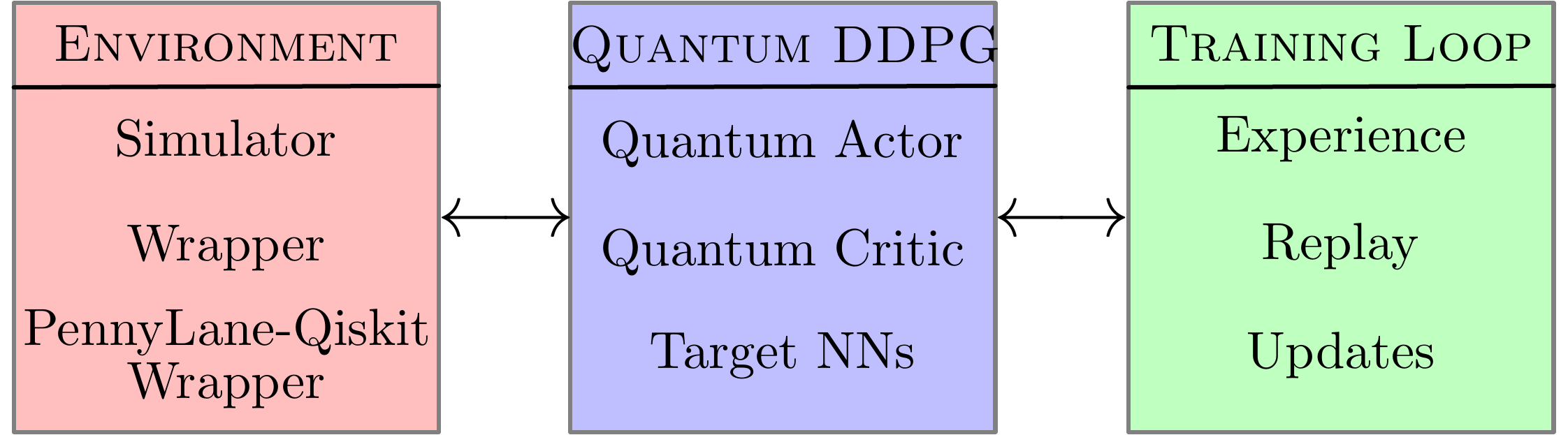}}
    \vspace{-.1cm}
    \caption{The building blocks of the proposed hybrid DRL framework.}
    \label{fig:methodology}
\end{figure}

These two challenges have motivated us to pursue a predictive scheme and a quantum algorithm in this paper. As shown in Fig. \ref{fig:methodology}, the DDPG agent uses predicted values to avoid the need to learn optimal actions, thereby enabling survivability under transient instability. The quantum algorithm simplifies the agent's training by estimating the actor and critic values in response to immediate actions. Now, let us delve into the operational aspects of the DRL agent alongside the proposed algorithm's functionalities. The observation/state ($s_t$) of the DDPG agent includes: generator/VSG rotor speeds ($\omega$), rotor angles ($\delta$), RoCoF ($\diff \omega/\diff t$), and active power outputs ($P_{e}$). The action ($a_t$) space comprises the power setpoint and inertia as $a_t = [\Delta P_{ref,1}, \dots, \Delta P_{ref,n}, \Delta J_1, \dots, \Delta J_n]$.

\subsection{Preliminaries of DDPG}

The DDPG algorithm is a class of model-free, off-policy actor-critic scheme that benefits from the architecture of both the deep Q network and the deterministic policy gradient. In this regard, the actor ($\mu(s|\theta^\mu)$) enacts to implement the deterministic policy that maximizes the rewards and consequently stimulates the critic ($Q(s,a|\theta^Q)$) to learn the value of state-action. Note that $\theta^\mu$ and $\theta^Q)$ are the parameters of the actor and critic networks, respectively. Hence, the actor network maps states to optimal deterministic continuous actions, and the critic network provides estimates of the value of state-action pairs accordingly. In addition, target actor ($\mu'(s|\theta^{\mu'})$) and critic ($Q'(s,a|\theta^{Q'})$) networks are integrated into the framework to increase the stability of the learning procedure of the DDPG agent. The objective of the actor network is to use the policy gradient theorem to maximize the expected $Q$-value as follows:
\begin{equation}
   \label{eq:eq20} 
   \nabla_{\theta^\mu} \mathbb{J} \approx \mathbb{E}_{s_t \sim \rho^\beta} \left[ \nabla_a Q(s,a|\theta^Q)|_{s=s_t,a=\mu(s_t)} \nabla_{\theta^\mu} \mu(s|\theta^\mu)|_{s=s_t} \right].
\end{equation}

Note that $\rho^\beta$ is the distribution of states under behavior policy $\beta$. Hence, the policy gradient theorem forces the actor's parameters to be updated based on the critic’s value for the chosen action. At the same time, the critic network aims to decrease the error of estimating the $Q$-value based on the target networks as follows
\begin{subequations}
\begin{align} 
 \mathbb{L}(\theta^Q) &= \mathbb{E}_{s_t,a_t,r_t,s_{t+1} \sim \mathcal{D}} \left[ \left( Q(s_t,a_t|\theta^Q) - y_t \right)^2 \right] \label{eq:eq21}\\
 y_t &= r_t + \gamma Q'(s_{t+1}, \mu'(s_{t+1}|\theta^{\mu'})|\theta^{Q'}). \label{eq:eq22} 
\end{align}
\end{subequations}

In this regard, $y_t$ is the target value obtained by the update process of the target networks based on the transition of experiences in the replay buffer, i.e., $(s_t,a_t,r_t,s_{t+1})$, as follows:

% \begin{equation}
% \label{eq:eq20}
%     |00\rangle = 1000 \quad |01\rangle = 0100 \quad |10\rangle = 0010\quad |11\rangle = 0001
% \end{equation}

\begin{align}
\label{eq:eq23}
\theta^{Q'} &\leftarrow \tau \theta^Q + (1-\tau)\theta^{Q'} \\
\theta^{\mu'} &\leftarrow \tau \theta^\mu + (1-\tau)\theta^{\mu'}
\end{align}

Note that $\tau \ll 1$ is suggested to be around $0.005$ for stability in the update process.

\subsection{Predictive DRL}
To increase the DRL agent's observability, the system state is augmented with predictions from a deep neural network (DNN). Hence, the new state vector is defined as $\tilde{s}_t = [s_t, \hat{\rho}_{t+\Delta T}]$. Here, $\hat{\rho}_{t+\Delta T}$ is the estimated value of the state variables prior to the onset of the transient instability. In this regard, the DNN architecture is trained on contingency scenarios to determine the stability status for a look-ahead time of $\Delta T$~\cite{masoumi2025transient}. In this study, the transient instability status (TIS), as defined subsequently, has been employed to train the DNN and assess how actions influence the rotor angle. Note that ${\delta _{\max }} = \max \{|{\delta_i} - {\delta_{j}}|\}, \ i,  j \in \{1,\dots,10\}$, is the maximum difference between the angles of any two generators. Then, we need to estimate $\widehat{TIS}_{t+\Delta T}$ to begin training our DDPG agent. Hence, the last step is to use an ensemble of DNN regressors ($n$) with prediction values (${\widehat{TIS} = } \frac{1}{N} \sum_{n=1}^N TIS_n$), standard deviations ($\sigma^2 = \frac{1}{N} \sum_{n=1}^N (TIS_n-\widehat{TIS})^2$) and confidence values ($C=1-\sigma$) to estimate $\widehat{TIS}_{t+\Delta T}$ based on the class of TIS. In doing so, the accumulated discounted reward function ($R_t = \sum_{k=0}^{T_e} \gamma^k r_{t+k}
$) is formulated to include the predictive terms as follows:

 % ${\text{TIS = }} \frac{{360 - {\delta _{\max }}}}{{360 + {\delta _{\max }}}}$

\begin{equation}
\label{eq:eq12}
{\text{TIS = }}\left\{ \begin{gathered}
  \text{class 1}{\text{, if}} \  \frac{{360 - {\delta _{\max }}}}{{360 + {\delta _{\max }}}} < 0 \hfill \\
  {\text{class 0, if}} \  \frac{{360 - {\delta _{\max }}}}{{360 + {\delta _{\max }}}} > 0 \hfill \\
\end{gathered}  \right.
\end{equation}

\begin{equation}
\label{eq:eq.13}
r_t = r_{\text{base}}(s_t, a_t) + r_{\text{pred}}(\hat{\rho}_{t+\Delta T})
\end{equation}
\begin{equation}
\label{eq:eq14}
\begin{aligned}
r_{\text{base}}(s_t, a_t)&=   - \xi_1 \Delta \omega(t)^2 - \xi_2 \cdot \left(\frac{\diff \omega}{\diff t}\Big|_t\right)^2 \\
& - \zeta \cdot \max_{t \in [t_0, t_f]} |\delta_{\max}(t) - \delta_{\text{COI}}(t)|  \\
& - \gamma_1 \sum_i \|(\Delta P_{\text{\text{ref}},i}(t)\\
&-\Delta P_{\text{\text{ref}},i}(t-1)) \|^2 \\
&- \gamma_2 \sum_i \|(\Delta J_{i}(t)-\Delta J_{i}(t-1))\|^2
\end{aligned}
\end{equation}
\begin{equation}\label{eq:eq15} 
r_{\text{pred}}(\hat{\rho}_{t+\Delta T}) = \begin{cases} -K_s\!\cdot\!\left(1 - \widehat{TIS}_{t+\Delta T}\right)^{2} & \begin{aligned}[t] &\text{if } \widehat{TIS}_{t+\Delta T} < \eta \\ &\text{and } C \geq C_{\text{th}} \end{aligned} \\[2pt] 10 & \text{otherwise.} 
\end{cases} 
\end{equation}
% \resizebox{\columnwidth}{!}{$
% r_{\text{pred}}(\hat{\rho}_{t+\Delta T}) =
% \begin{cases}
% -K_s\bigl(1-\widehat{TIS}_{t+\Delta T}\bigr)^2
% & \text{if } \widehat{TIS}_{t+\Delta T}<\eta \text{ and } C\ge C_{\text{th}} \\[2pt]
% 10 & \text{otherwise.}
% \end{cases}
% $}

% \begin{equation}
% \label{eq:eq15}
% r_{\text{pred}}(\hat{\rho}_{t+\Delta T}) =
% \begin{cases}
% -K_s \cdot \left(1 - \widehat{TIS}_{t+\Delta T}\right)^2 & \text{if } \quad \widehat{TIS}_{t+\Delta T} < \eta   \& C \geq C_{\text{th}} \\
% 10 & \text{otherwise}
% \end{cases}
% \end{equation}

In our predictive framework, $\eta (0.1)$, $C_{\text{th}} (0.95)$, and $K_s$ are the safety threshold, the confidence threshold of the estimation, and the constant penalty terms that dominate the reward signal. Moreover, the parameters of the reward function are predefined as $\xi_1$, $\xi_2$, $\zeta$, $\gamma_1$, $\gamma_2$, and $K_s$ to be 100, 0.1, 0.01, 0.001, 0.001, and 100, respectively. Note that $T_e$ is the maximum number of time steps in each episode.

% \begin{equation}
% \label{14}
%   r_{\text{base}}(s_t, a_t) = - \left( \alpha_1 \cdot \Delta \omega_t^2 + \alpha_2 \cdot \left(\frac{d\omega}{dt}\Big|_t\right)^2 + (\frac{d\omega}{dt}\right)^2 \\ 
%   + \beta \cdot \max_{t \in [t_0, t_f]} |\delta_{\max}(t) - \delta_{\text{COI}}(t)|   
% \end{equation}

% \begin{align}
% \label{eq:eq13}
% {\widehat{TIS} = } \frac{1}{N} \sum_{n=1}^N TIS_n\\
% \label{eq:eq14}
% \sigma^2 = \frac{1}{N} \sum_{n=1}^N (TIS_n-\widehat{TIS})^2\\
% \label{eq:eq15}
% C=1-\sigma
% \end{align}

% \begin{align}
% \label{eq:eq14}
% \sigma^2 = \frac{1}{N} \sum_{n=1}^N (TIS_n-\widehat{TIS})^2
% \end{align}
% \begin{align}
% \label{eq:eq15}
% C=1-\sigma
% \end{align}

\subsection{Preliminaries of Quantum Computing}
Classical computing can only be performed in two binary states: $0$ and $1$. However, quantum computing can represent the information units in the superposition of $\ket{0}$ and $\ket{1}$. The reason is the ability to trace the energy levels of excited states in quantum mechanics. This is considered the first advantage of using quantum computing to analyze multiple solution combinations. Typically, a qubit can be defined based on two distinct energy levels in the Z quantum (Pauli) space
\begin{equation}
\label{eq:eq18}
    \ket{\psi} = \alpha \ket{0} + \beta \ket{1},
\end{equation}
\noindent where $\ket{\psi}$, $\alpha$, and $\beta$ are quantum information and probability amplitudes of Z-Pauli measurement for being in states of $\ket{0}$ and $\ket{1}$, respectively. The second advantage is the use of entanglement to create a multi-qubit information system based on the interactions of single qubits. On the other hand, the transformation of the two-qubit state into the classical state can be represented as 
\begin{equation}
\label{eq:eq19}
    \ket{00} = 1000 \quad \ket{01} = 0100 \quad \ket{10} = 0010\quad \ket{11} = 0001.
\end{equation}
Thus, the key insight lies in substituting $N$ classical information with $n$ units of quantum information, where $n = \lceil \log_2 N \rceil$. This substitution serves as the impetus for adopting the quantum approach described herein, specifically to reduce the computational demands associated with complex power system dynamics during the inference stage of the proposed DDPG. 
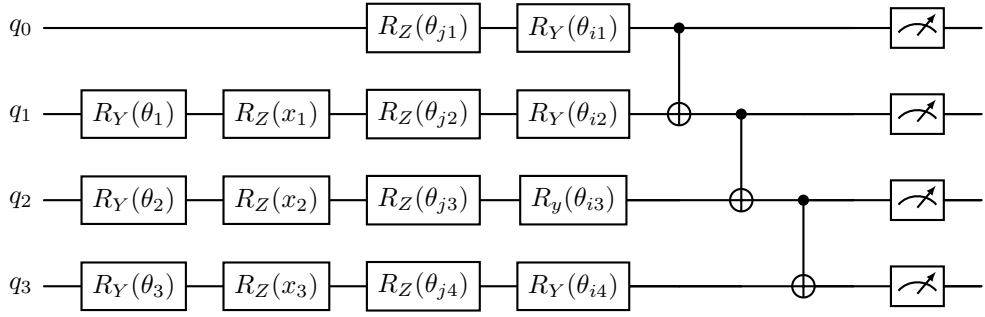
\begin{figure*}[htbp]
    \centering
    % \begin{adjustbox}{maxwidth=\columnwidth}
    \begin{quantikz}
        \lstick{$q_0$}  & & &  \gate{R_Z(\theta_{j1})}  & \gate{R_Y(\theta_{i1})} &\ctrl{1} & \qw      & \qw      & \qw      & \meter{} & \qw \\
        \lstick{$q_1$}  & \gate{R_Y(\theta_1)} & \gate{R_Z(x_1)} & \gate{R_Z(\theta_{j2})} & \gate{R_Y(\theta_{i2})} & \targ{} & \ctrl{1} & \qw      & \qw      & \meter{} & \qw \\
        \lstick{$q_2$}  & \gate{R_Y(\theta_2)} & \gate{R_Z(x_2)} & \gate{R_Z(\theta_{j3})} & \gate{R_y(\theta_{i3})} & \qw     & \targ{}  & \ctrl{1} & \qw      & \meter{} & \qw \\
        \lstick{$q_3$} & \gate{R_Y(\theta_3)} & \gate{R_Z(x_3)} & \gate{R_Z(\theta_{j4})}  & \gate{R_Y(\theta_{i4})}  &\qw     & \qw      & \targ{}  & \qw      &\meter{} & \qw \\
    \end{quantikz}
    % \end{adjustbox}
    \vspace{-.2in}
    \caption{The four-qubit ansatz for three observations and one action.}
    \label{fig:ansatz}
\end{figure*}

\subsection{Application of PQC in DDPG}
As mentioned, superposition and entanglement are the stepping stones of quantum mechanics~\cite{chen2026quantum}. In this regard, superposition aims to generate combinations of complex states based on the problem's dimensionality. Moreover, entanglement has the duty of creating the possibility of a solution based on the connection of the qubits~\cite{dragan2025continuous}. In the quantum machine learning application, we must consider another step before encoding the data from the classical representations to the quantum basis. In addition, $R_Y$, $R_Z$, and $CNOT$ are utilized as operators of the encoding and entanglement. In other words, they are trainable single-qubit rotations~\cite{lan2021variational}. It is worth noting that Pauli-Z and Pauli-X are measurement operators used in the actor and critic networks, respectively. Hence, the unitary layer can be formulated as a hierarchy of applying the gates and entangling the tensor product. Note that the hierarchy is set up so that input rotations are first, then trainable single-qubit rotations, and then entanglement as
\begin{equation}\label{20}
\begin{aligned}
U_\ell(x;\alpha,\beta,\theta)
&= E \circ
\Bigg(\bigotimes_{j=0}^{n-1} R_Y(\theta_{\ell,j}^{(2)})\,R_Z(\theta_{\ell,j}^{(1)})\Bigg) \\
&\quad \circ\;
\Bigg(\bigotimes_{j=0}^{m-1} R_Z(\beta_{\ell,j} x_j)\, R_Y(\alpha_{\ell,j} x_j)\Bigg),
\end{aligned}
\end{equation}
% \begin{equation}
% \label{20}
%     U_\ell(x;\alpha,\beta,\theta)
% =
% E
% \;\circ\;
% \Bigg(\bigotimes_{j=0}^{n-1} R_Y(\theta_{\ell,j}^{(2)})R_Z(\theta_{\ell,j}^{(1)})\Bigg)
% \;\circ\;
% \Bigg(\bigotimes_{j=0}^{m-1} R_Z(\beta_{\ell,j}\, x_j)\,R_Y(\alpha_{\ell,j}\, x_j)\Bigg),
% \end{equation}
\noindent where $x$, $\alpha$, and $\beta$ are classical input vectors and input scaling parameters per layer. In addition, $\theta$ and $E$ are trainable parameters and entangler ($CNOT$), respectively. Hence, the full circuit operation after $L$ layers is formulated as
\begin{equation}
    \label{21}
\ket{\psi} = U_L\,U_{L-1}\cdots U_1\,\ket{\mathbf{0}}.
\end{equation}

Hence, the quantum circuit implements a unitary transformation as
\begin{equation}
\label{22}
U(\theta) = \prod_{l=1}^{ L} U_{\text{ent}} U_{\text{rot}}(\theta_l) U_{\text{encode}}(x),
\end{equation}
\noindent where $U_{\text{encode}}(x)$, $U_{\text{rot}}(\theta_l)$, and $U_{\text{ent}}$ are applied to encode classical data into quantum states, build parameterized rotations, and create entanglement between qubits throughout Layers $L$.

\subsection{Actor Network}
The justification for ansatz selection of the actor network is based on three characteristics. The first is expressibility through combinations of $R_Y(\theta)$, $R_Z(\theta)$, and $CNOT$ gates. The ansatz must be trainable to provide levels of information for the measurement step. Hence, the quantum actor network $\mu_Q(s|\theta^\mu)$ is defined with respect to the trainable parameters $\theta$ as
\begin{equation}
\label{23}
\mu_Q(s) = f_{\text{post}}(  \bra{0} U^\dagger(s, \theta^\mu) Z_i U(s, \theta^\mu) \ket{0} ),
\end{equation}
where $U(s, \theta^\mu)$ and $f_{\text{post}}$ are the applied ansatz and the classical post-processing function to determine the action in response to the states of the environment, respectively, and $U^\dagger$ denotes the complex conjugate of $U$. In addition, $Z_i$ are the Pauli-Z measurement operators applied to all qubits. In this regard, the quantum actor implements the mapping via $f_{\text{post}}$ as
\begin{equation}
\label{24}
\mu_Q(s) = A \cdot \tanh\left( \frac{1}{K} \sum_{i=1}^{N_a} \braket{Z_i} \right) + B,
\end{equation}
\noindent where $A$ and $B$ are the scale of action states and bias vectors, respectively. In addition, $N_a$, $K$, and $\braket{Z_i}$ are the number of actions, normalization constant, and expectation values from quantum measurements, respectively. Now that we have described the PQC step of the actor network, the only task is to compute the gradient of the training and update the actor using the parameter-shift rule as follows:
\begin{subequations}\label{eq:25}
\begin{align}
\frac{\partial \mu_Q(s)}{\partial \theta_i}
&= \tfrac{1}{2}\!\left[\mu_Q\!\left(s;\,\theta_i + \tfrac{\pi}{2}\right)
      - \mu_Q\!\left(s;\,\theta_i - \tfrac{\pi}{2}\right)\right]
\label{eq:25a} \\[4pt]
\nabla_{\theta^\mu} \mathbb{J}
&\approx \mathbb{E}_{s \sim \mathcal{D}}
\!\left[\, \nabla_a Q_Q\!\left(s,a \mid \theta^Q\right)\big|_{a=\mu_Q(s)}\;
          \nabla_{\theta^\mu}\mu_Q\!\left(s \mid \theta^\mu\right) \right]
\label{eq:25b}
\end{align}
\end{subequations}

% \begin{equation}
% \label{25}
% \frac{\partial \mu_Q(s)}{\partial \theta_i} = \frac{1}{2} \left[ \mu_Q(s; \theta_i + \frac{\pi}{2}) - \mu_Q(s; \theta_i - \frac{\pi}{2}) \right]
% \end{equation}

% \begin{equation}
% \nabla_{\theta^\mu} \mathbb{J} \approx \mathbb{E}_{s \sim \mathcal{D}} \left[ \nabla_a Q_Q(s,a|\theta^Q)|_{a=\mu_Q(s)} \nabla_{\theta^\mu} \mu_Q(s|\theta^\mu) \right]
% \end{equation}

\subsection{Critic Network}
The ansatz of the critic network $Q_Q(s,a|\theta^Q)$ is described as
\begin{equation}
\label{26}
Q_Q(s,a) = w_{\text{out}} \cdot \bra{0} U^\dagger([s,a], \theta^Q) X_0 U([s,a], \theta^Q) \ket{0} + b_{\text{out}},
\end{equation}

\noindent where $w_{\text{out}}$ and $b_{\text{out}}$ express the trainable output scaling parameters. In addition, $[s,a]$ is used to represent the concatenation of state and action. Here, the observable is the Pauli-X measurement on the first qubit, which captures the Q-value of the state-action pair. In other words, the critic ansatz stacks the single observation of the expected value of the first qubit into a tensor. This value is multiplied by the output scaling parameters. In doing so, the critic update is rewritten as
% \begin{equation}\label{27}
% \mathbb{L}(\theta^Q) = \mathbb{E}_{(s,a,r,s') \sim \mathcal{D}} \left[ \left( Q_Q(s,a|\theta^Q) - (r + \gamma Q'_Q(s', \mu'_Q(s'|\theta^{\mu'})|\theta^{Q'}) \right)^2 \right]
% \end{equation}
\begin{equation}\label{27}
\begin{aligned}
\mathbb{L}(\theta^Q)
&= \mathbb{E}_{(s,a,r,s') \sim \mathcal{D}} \Bigg[ \Big(
    Q_Q(s,a\mid\theta^Q) \\
&\qquad\qquad - \big(r + \gamma\, Q'_Q\!\big(s',\, \mu'_Q(s'\mid\theta^{\mu'}) \mid \theta^{Q'}\big)\big)
\Big)^2 \Bigg].
\end{aligned}
\end{equation}

\section{Case Study Description and Implementation}
This study evaluates the quantum-enhanced frequency response within the modified IEEE 39-bus test system, where $50\%$ of the total generation is provided by IBRs. A three-phase short-circuit event was applied at 95\% of Line 33, in proximity to Bus 26, occurring at $t = 2$ s. The frequency incident (load increase) is set to $0.6$ pu (Bus 31). Then, we monitor the dynamics for 5 s. It should be noted that TDS, converter dynamics (REGCV2), and scenario generation were derived using the ANDES library~\cite{cui2025andes}. For configuring the DDPG, Python 3 along with the Tianshou library~\cite{weng2022tianshou} were used. The quantum framework of the PQC is implemented through the PennyLane library~\cite{bergholm2018pennylane}. Furthermore, the actor's PQC comprises 20 qubits, achieved by substituting 5 SGs with IBR, and spans four observation spaces, whereas the critic incorporates 30 qubits for state-action pair assessment. To address this constraint, Fig.~\ref{fig:ansatz} illustrates an instance featuring 3 observations and a single action. As indicated, the data encoding phase of the proposed algorithm involves applying the initial two rotation gates, with subsequent gates enhancing the circuit's expressibility. In addition, the circuit's depth is determined as 3 layers through experimental optimization.
% It should be noted that the TDS and the scenario generation were obtained using the ANDES library~\cite{cui2025andes}. Python 3 and the Tianshou~\cite{weng2022tianshou} library have been used to parameterize the DDPG. The quantum architecture of the PQC is utilized based on the PennyLane library ~\cite{bergholm2018pennylane}. In addition, the PQC of the actor has 20 qubits, resulting from replacing 5 SGs with IBR, and has four observation spaces. This is while the critic has 30 qubits due to the evaluation of the state-action pair. To ease this limitation, Fig. \ref{fig:ansatz} shows an example of 3 observations with only one action. As shown, the first two rotation gates are applied during the data encoding step of the proposed algorithm. The remaining rotation gates are then used to increase the circuit's expressibility. Also, the depth of the circuit is set to 3 layers based on trial and error.    

\section{Results and Discussions}

This section represents the results based on the performance of the DDPG agent throughout the episodes. The agent enacts after the fault clearance and  aims to revive the rotor angle stability based on the defined reward function. The maximum number of time steps in each episode is 100. As mentioned, the hybrid approach is initiated based on the performance of the DNN architecture to estimate the status of the contingency. According to Table~\ref {Table1}, DNN prevails over both machine learning and DL counterparts, namely Bayesian regularization (BR) and convolutional neural networks (CNN), respectively. Hence, the dependency of the conducted DNN is proven accordingly. 

\vspace{-.2cm}
\begin{table}[h!]
\caption{Comparison of Accuracy in the TIS Estimation}
% \vspace{-.5cm}
\label{Table1}
\begin{center}
\begin{tabular}{c|c|c|c}
\hline
\hline
 & CNN~\cite{gupta2018online}  & BR~\cite{zare2019intelligent}& DNN~\cite{masoumi2025transient}\\
\hline
\textbf{Accuracy} [$\%$] & $89.22$ & $91.6$ & $\mathbf{99}$\\
\hline
\hline
\end{tabular}
\end{center}
\vspace{-.2cm}
\end{table}

\begin{figure}[htbp]
\centerline{\includegraphics[width=\linewidth]{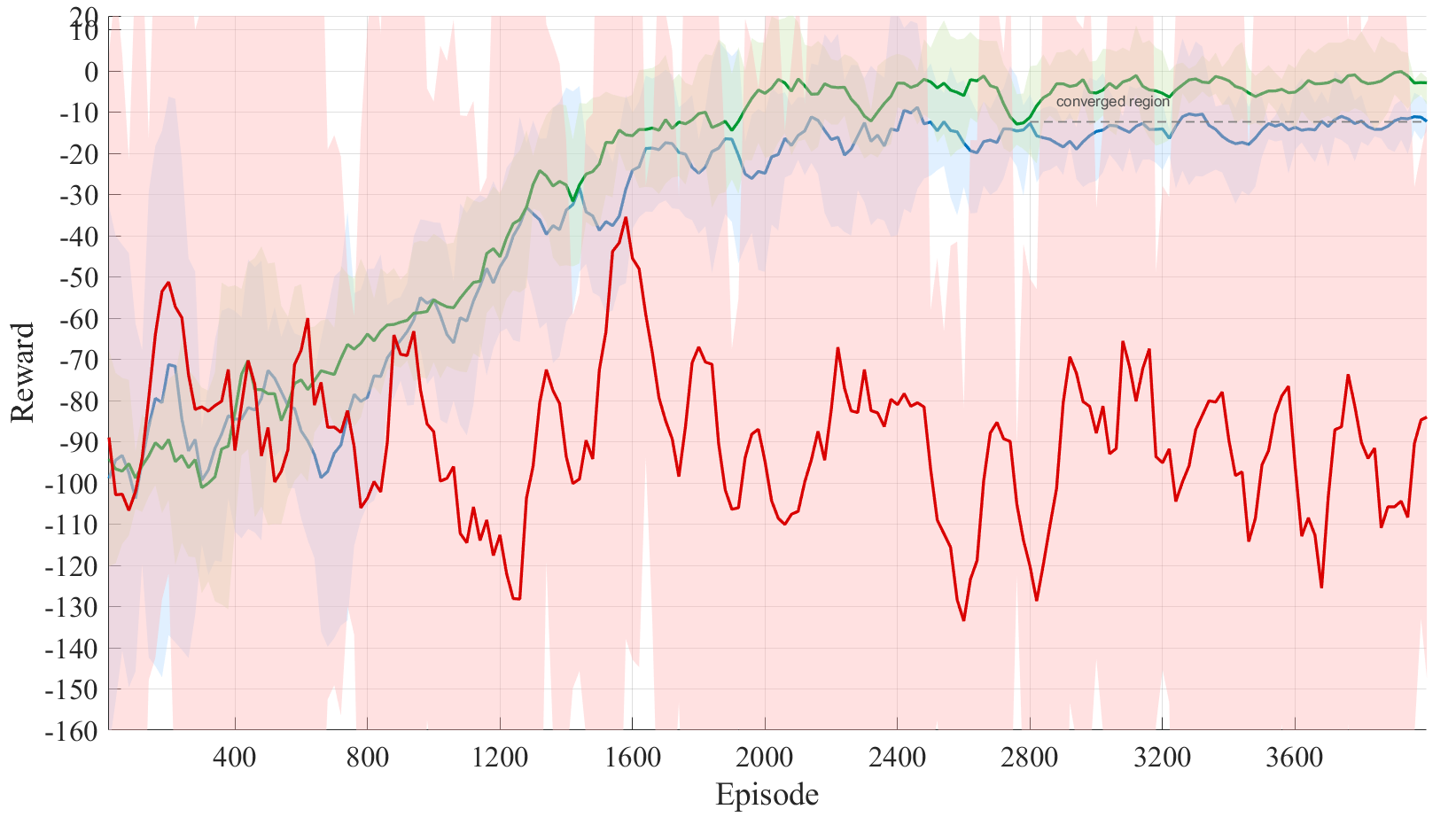}}
    \vspace{-.1cm}
    \caption{The reward function: the \textcolor{Blue}{blue}, \textcolor{Green}{green}, and \textcolor{Crimson}{red} curves indicate the quantum-embedded DDPG agent, vanilla DDPG agent, and the performance under noise. Also, the shaded areas show the standard deviation from the conducted simulation over 10 consecutive runs.}
    \label{fig:Reward_fucntion}
\end{figure}

We also examined the DRL agent's ability to discern environmental response patterns with a reduced set of trainable parameters. In this context, the traditional approach encapsulates the critic network's weights and biases, incorporating $30 \times 128 \times 128 \times 128$ trainable parameters. In this context, the quantum-embedded algorithm uses 100 adjustable parameters per quantum layer. Figure 3 illustrates the quantum-embedded DDPG agent's performance. As depicted, the behavior of the quantum-embedded system aligns with that of the conventional agent. Note that the solid lines correspond to the reward function as computed from (\ref{eq:eq.13}).
% The blue, green, and red curves in Fig.~\ref{fig:Reward_fucntion} indicate the quantum-embedded DDPG agent, vanilla DDPG agent, and the performance under noise. 
From Fig.~\ref{fig:Reward_fucntion}, it is evident that the quantum-enhanced DDPG agent closely approximates the performance of its predecessor, effectively aiming to reduce the effects of dynamic security contingencies, as indicated by the reward function. The observed trend correlates with the standard deviation. This suggests that the proposed algorithm is highly similar to the vanilla DDPG agent, consistent with the objectives of this study. However, the classical DDPG agent outperforms the proposed algorithm due to its inherent ability to handle greater complexity, while the proposed algorithm meets DSC requirements with significantly fewer parameters. Another consideration is the influence of noise in the noisy intermediate-scale quantum (NISQ) era. In realistic scenarios, the algorithm’s performance under significant noise fails to satisfy even the basic DSC regulatory needs, as illustrated by the \textcolor{Crimson}{red} curve. The NISQ model's standard deviation reveals uncertainty about noise effects in the near-term\footnote{``Near-term'' generally applies to algorithms and codes running on quantum hardware; please refer to \cite{benchmarking2024pennylane} for one application.} applications. Moreover, Fig. 4 demonstrates the proposed method's performance with respect to rotor angle deviation. The data show that rotor angles are notably affected by severe contingencies, rendering the studied system vulnerable to cascading failures if the contingency exceeds CCT. Therefore, the proposed algorithm successfully mitigates the contingency impact, restoring rotor angles to stable operation boundaries, thereby confirming the satisfactory efficacy of the implemented DSC regularization strategy.

\begin{figure}[htbp]
\centerline{\includegraphics[width=\linewidth]{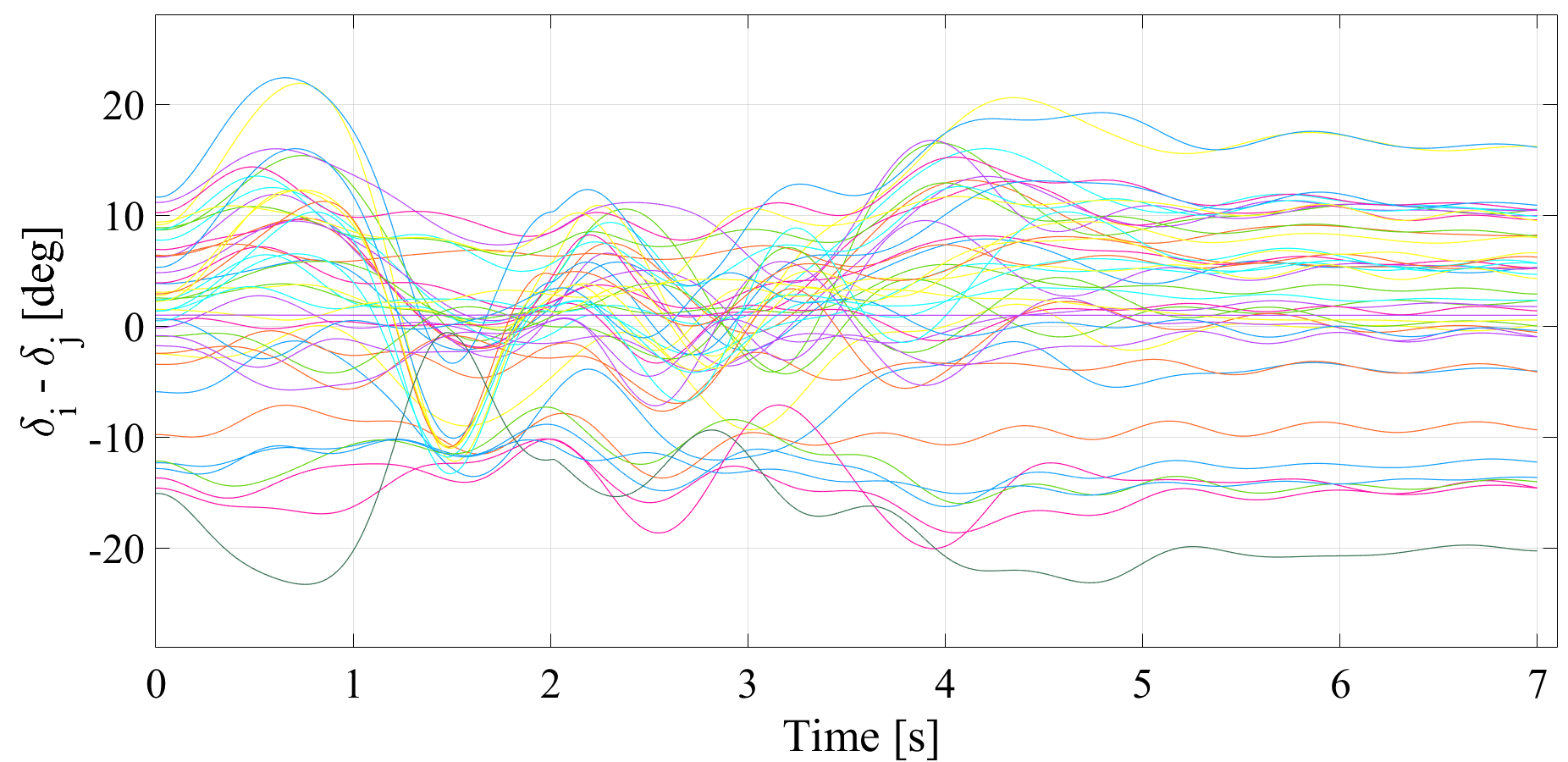}}
    \vspace{-.1cm}
    \caption{The rotor angle difference of the test system.}
    \label{fig:fig2}
\end{figure}

% \ssection{Training in Quantum Simulator}

% \subsection{Load Increase under OOS}

% \subsection{Performance Evaluation and Noise Correction}

\section{Conclusion and Future Directions}

This paper presents a comprehensive view of the application of quantum computing in the DSC of IBR-rich grids. In particular, the proposed methodology develops PQCs to model the actor and critic networks of the vanilla DDPG agent. The simulation results demonstrate the validity of the designed ansatz by indicating its compatibility with its predecessor in reviving the system dynamics to a stable state after a severe rotor angle instability incident. Empirical evidence suggests that the quantum-integrated method encounters constraints when aiming for optimal precision in practical DSC applications. Consequently, quantum algorithms utilizing PQCs require enhanced modeling to account for noise, which is achieved by integrating specific rotation gates during the state preparation and measurement (SPAM) phases. This approach mimics the detrimental effects of errors characteristic of the NISQ era. Thus, meticulous care and comprehensive modeling during the quantum state transpilation phase are indispensable.

\bibliographystyle{IEEEtran}
\bibliography{IEEEabrv,References}

\begin{thebibliography}{10}
\providecommand{\url}[1]{#1}
\csname url@samestyle\endcsname
\providecommand{\newblock}{\relax}
\providecommand{\bibinfo}[2]{#2}
\providecommand{\BIBentrySTDinterwordspacing}{\spaceskip=0pt\relax}
\providecommand{\BIBentryALTinterwordstretchfactor}{4}
\providecommand{\BIBentryALTinterwordspacing}{\spaceskip=\fontdimen2\font plus
\BIBentryALTinterwordstretchfactor\fontdimen3\font minus \fontdimen4\font\relax}
\providecommand{\BIBforeignlanguage}[2]{{%
\expandafter\ifx\csname l@#1\endcsname\relax
\typeout{** WARNING: IEEEtran.bst: No hyphenation pattern has been}%
\typeout{** loaded for the language `#1'. Using the pattern for}%
\typeout{** the default language instead.}%
\else
\language=\csname l@#1\endcsname
\fi
#2}}
\providecommand{\BIBdecl}{\relax}
\BIBdecl

\bibitem{bevrani2017microgrid}
H.~Bevrani, B.~Fran{\c{c}}ois, and T.~Ise, \emph{Microgrid Dynamics and Control}.\hskip 1em plus 0.5em minus 0.4em\relax John Wiley \& Sons, 2017.

\bibitem{shazon2022frequency}
M.~N.~H. Shazon, A.~Jawad \emph{et~al.}, ``Frequency control challenges and potential countermeasures in future low-inertia power systems: A review,'' \emph{Energy Reports}, vol.~8, pp. 6191--6219, 2022.

\bibitem{wang2025transient}
L.~Wang, P.~Hu, and X.~Wang, ``Transient stability analysis of grid-forming {VSC} based on hybrid synchronization control under asymmetrical {AC} faults,'' \emph{Sustain. Energy Technol. Assess.}, vol.~82, p. 104534, Oct. 2025.

\bibitem{jafari2024role}
M.~Jafari, G.~B. Gharehpetian, and A.~Anvari-Moghaddam, ``On the role of virtual inertia units in modern power systems: A review of control strategies, applications and recent developments,'' \emph{Int. J. Electr. Power Energy Syst.}, vol. 159, p. 110067, Aug. 2024.

\bibitem{fang2025dynamic}
C.~Fang \emph{et~al.}, ``Dynamic interaction analysis of power systems connected with grid-forming converters,'' \emph{Int. J. Electr. Power Energy Syst.}, vol. 171, p. 110908, Oct. 2025.

\bibitem{saleem2024assessment}
M.~I. Saleem and S.~Saha, ``Assessment of frequency stability and required inertial support for power grids with high penetration of renewable energy sources,'' \emph{Electr. Power Syst. Res.}, vol. 229, p. 110184, Apr. 2024.

\bibitem{zhang2021grid}
H.~Zhang, W.~Xiang, W.~Lin, and J.~Wen, ``Grid forming converters in renewable energy sources dominated power grid: Control strategy, stability, application, and challenges,'' \emph{J. Mod. Power Syst. Clean Energy}, vol.~9, no.~6, pp. 1239--1256, Nov. 2021.

\bibitem{abdelghany2025enhanced}
M.~B. Abdelghany \emph{et~al.}, ``Enhanced grid-forming operation of virtual synchronous generator units,'' \emph{{IEEE} Trans. Power Del.}, 2025, early access.

\bibitem{saha2023impact}
S.~Saha, M.~I. Saleem, and T.~K. Roy, ``Impact of high penetration of renewable energy sources on grid frequency behaviour,'' \emph{Int. J. Electr. Power Energy Syst.}, vol. 145, p. 108701, 2023.

\bibitem{ma2025coordinated}
N.~Ma, F.~Bai, and T.~K. Saha, ``Coordinated frequency control based on hierarchical {MPC} considering tie-line power flow,'' \emph{{IEEE} Trans. Power Syst.}, 2025, early access.

\bibitem{chen2021enhanced}
M.~Chen, D.~Zhou, and F.~Blaabjerg, ``Enhanced transient angle stability control of grid-forming converter based on virtual synchronous generator,'' \emph{{IEEE} Trans. Ind. Electron.}, vol.~69, no.~9, pp. 9133--9144, Sep. 2022.

\bibitem{eskandari2023deep}
M.~Eskandari, A.~V. Savkin, and J.~Fletcher, ``A deep reinforcement learning-based intelligent grid-forming inverter for inertia synthesis by impedance emulation,'' \emph{{IEEE} Trans. Power Syst.}, vol.~38, no.~3, pp. 2978--2981, May 2023.

\bibitem{oboreh2023virtual}
O.~Oboreh-Snapps \emph{et~al.}, ``Virtual synchronous generator control using twin delayed deep deterministic policy gradient method,'' \emph{{IEEE} Trans. Energy Convers.}, vol.~39, no.~1, pp. 214--228, Mar. 2024.

\bibitem{lee2024deep}
W.-G. Lee and H.-M. Kim, ``Deep reinforcement learning-based dynamic droop control strategy for real-time optimal operation and frequency regulation,'' \emph{IEEE Trans. Sustain. Energy}, vol.~16, no.~1, pp. 284--294, Jan. 2025.

\bibitem{chen2023investigation}
L.~Chen \emph{et~al.}, ``Investigation on transient stability enhancement of multi-{VSG} system incorporating resistive {SFCL}s based on deep reinforcement learning,'' \emph{{IEEE} Trans. Ind. Appl.}, vol.~60, no.~1, pp. 1780--1793, Jan.--Feb. 2024.

\bibitem{yang2022distributed}
Q.~Yang, L.~Yan, X.~Chen, Y.~Chen, and J.~Wen, ``A distributed dynamic inertia-droop control strategy based on multi-agent deep reinforcement learning for multiple paralleled {VSG}s,'' \emph{{IEEE} Trans. Power Syst.}, vol.~38, no.~6, pp. 5598--5612, Nov. 2023.

\bibitem{eskandari2023convolutional}
M.~Eskandari, A.~V. Savkin, and J.~Fletcher, ``Convolutional neural network with reinforcement learning for trajectories boundedness of fault ride-through transients of grid-feeding converters in microgrids,'' \emph{{IEEE} Trans. Ind. Informat.}, vol.~20, no.~3, pp. 4906--4918, Mar. 2024.

\bibitem{ma2024dynamic}
Y.~Ma, Z.~Hu, and Y.~Song, ``Dynamic nonlinear droop-based fast frequency regulation for power systems with flexible resources using meta-reinforcement learning approach,'' \emph{J. Mod. Power Syst. Clean Energy}, vol.~13, no.~2, pp. 379--390, Mar. 2025.

\bibitem{lockwood2020reinforcement}
O.~Lockwood and M.~Si, ``Reinforcement learning with quantum variational circuit,'' in \emph{Proc. AAAI Conf. Artif. Intell. Interact. Digital Entertain.}, vol.~16, no.~1, 2020, pp. 245--251.

\bibitem{chen2024deep}
H.-Y. Chen, Y.-J. Chang, S.-W. Liao, and C.-R. Chang, ``Deep {Q}-learning with hybrid quantum neural network on solving maze problems,'' \emph{Quantum Mach. Intell.}, vol.~6, no.~1, p.~2, 2024.

\bibitem{wu2025quantum}
S.~Wu, S.~Jin, D.~Wen, D.~Han, and X.~Wang, ``Quantum reinforcement learning in continuous action space,'' \emph{Quantum}, vol.~9, p. 1660, 2025.

\bibitem{li2025power}
J.~Li, R.~Shi, Z.~Dong, J.~Li, and X.~Zhang, ``Power-frequency oscillation modeling, analysis and suppression of multi-{VSG} grid-connected system,'' \emph{Int. J. Electr. Power Energy Syst.}, vol. 172, p. 111129, Nov. 2025.

\bibitem{masoumi2025transient}
A.~Masoumi and M.~Korkali, ``Transient-stability-aware frequency provision in {IBR}-rich grids via information gap decision theory and deep learning,'' \emph{arXiv preprint arXiv:2507.13265}, 2025.

\bibitem{chen2026quantum}
S.~Y.-C. Chen, ``Quantum reinforcement learning: Concepts and applications,'' \emph{Quantum Comput. AI}, pp. 3--23, 2026.

\bibitem{dragan2025continuous}
T.-A. Dragan, A.~K{\"u}nzner, R.~Wille, and J.~M. Lorenz, ``Continuous quantum reinforcement learning for robot navigation,'' in \emph{Proc. Int. Conf. Agents Artif. Intell.}, 2025.

\bibitem{lan2021variational}
Q.~Lan, ``Variational quantum soft actor-critic,'' \emph{arXiv preprint arXiv:2112.11921}, 2021.

\bibitem{cui2025andes}
H.~Cui, \emph{ANDES Manual ({R}elease 1.9.3.post9+g243a7da)}, May 2025.

\bibitem{weng2022tianshou}
J.~Weng \emph{et~al.}, ``Tianshou: A highly modularized deep reinforcement learning library,'' \emph{J. Mach. Learn. Res.}, vol.~23, no. 267, pp. 1--6, 2022.

\bibitem{bergholm2018pennylane}
V.~Bergholm \emph{et~al.}, ``Pennylane: Automatic differentiation of hybrid quantum-classical computations,'' \emph{arXiv preprint arXiv:1811.04968}, 2018.

\bibitem{gupta2018online}
A.~Gupta, G.~Gurrala, and P.~S. Sastry, ``An online power system stability monitoring system using convolutional neural networks,'' \emph{{IEEE} Trans. Power Syst.}, vol.~34, no.~2, pp. 864--872, Mar. 2019.

\bibitem{zare2019intelligent}
H.~Zare, Y.~Alinejad-Beromi, and H.~Yaghobi, ``Intelligent prediction of out-of-step condition on synchronous generators because of transient instability crisis,'' \emph{Int. Trans. Electr. Energy Syst.}, vol.~29, no.~1, p. e2686, 2019.

\bibitem{benchmarking2024pennylane}
{Benchmarking Near-term Quantum Algorithms is an Art}. \url{https://pennylane.ai/blog/2024/03/benchmarking-near-term-quantum-algorithms-is-an-art}. Accessed on 10/15/2025.

\end{thebibliography}

% The 'triggered' command can be changed if desired:


%\bibliographystyle{IEEEtran}
% \bibitem{Shell}
% M.~Shell, \emph{How to Use the IEEEtran Latex Class}, Latex Archive Contents, \verb+http://www.ieee.org/conferences_events/+ \verb+conferences/publishing/templates.htm+


% \end{thebibliography}

\end{document}